\documentclass[letterpaper]{article}
\usepackage{kotex}
\usepackage[utf8]{inputenc}
\usepackage{titlesec}
\usepackage{isea}
\usepackage[pdftex]{graphicx}
\usepackage{times}
\usepackage{helvet}
\usepackage{courier}
\usepackage[numbers]{natbib}
\usepackage{url}
\usepackage{hyperref}
\title{Immersive Fantasy Based on Digital Nostalgia : Environmental Narratives for the Korean Millennials and Gen Z}
\author{
Yerin Doh\textsuperscript{1}\thanks{These authors contributed equally as co-first authors.},
Joonhyung Bae\textsuperscript{2}\footnotemark[1]\\
\\
\textsuperscript{1}Independent Artist, Seoul, Republic of Korea, \texttt{luxembur8er@gmail.com}\\
\textsuperscript{2}Graduate School of Culture Technology, KAIST, Daejeon, Republic of Korea, \texttt{jh.bae@kaist.ac.kr}
}

\begin{document} 
\maketitle
\begin{abstract}
This study introduces the media artwork \textit{Dear Passenger, Please Wear a Mask}, designed to offer a layered exploration of single-use mask waste, which escalated during the COVID-19 pandemic. The piece reframes underappreciated ecological concerns by interweaving digital nostalgia and airline travel recollections of Millennials and Gen Z with a unique fantasy narrative. Via a point-and-click game and an immersive exhibition, participants traverse both virtual and real domains, facing ethical and environmental dilemmas. While it fosters empathy and potential action, resource use and post-experience engagement challenges persist.
\end{abstract}


Digital Nostalgia, Interactive Art, Environmental Narratives, Fantasy Storytelling, Cultural Memory

\section{Introduction}

Climate change and environmental pollution are complex issues that demand not only institutional or technological solutions but also ethical and cultural engagement \cite{Kumar}. The recent prominence of “eco-anxiety” underscores the broad emotional and societal impact of the climate crisis \cite{Pihkala2020}, suggesting that art can foster empathy and inspire action through sensory experiences \cite{annurev:/content/journals/10.1146/annurev.environ.31.042605.134920}.

Contemporary artists have employed physical materials—such as glaciers and solar energy—and digital forms like augmented reality (AR) and non-fungible tokens (NFTs) to visualize climate change and resource depletion from multiple perspectives \cite{BusinessWire2023,ZHANG2021594}. Nevertheless, the so-called “environmental paradox,” arising from high energy consumption, remains a concern. In South Korea, diverse initiatives—including upcycling projects and immersive exhibitions—are underway \cite{Chae2021,Ryu2024}. In particular, Millennials and Gen Z (born from the early 1980s to the mid-2000s; hereafter, “MZ”) stand out for their digital sensibilities and nostalgia culture, opening new avenues for environmental art.

This study proposes and analyzes a media artwork that integrates the MZ’s early digital experiences (e.g., browser games) with fantasy narratives. It aims to explore how digital nostalgia can elevate public awareness of environmental problems and motivate behavioral change.

\section{Related Works}

\subsection{The MZ and Digital Nostalgia}

Following the liberalization of overseas travel in Korea \cite{ART003060363}, members of the MZ grew accustomed to flying despite the environmental dilemma of significant greenhouse gas emissions from air travel \cite{IPCC1999}. They also developed a shared digital nostalgia through browser games played in PC cafés or school computer labs, paralleling global web-based games such as Neopets or Club Penguin \cite{Fox2021}. This nostalgia can reinforce communal bonds and personal continuity \cite{ZHANG2021594}, potentially leveraging memories to heighten awareness of present climate crises \cite{Lee2021}.

\subsection{Fantasy Narratives and Environmental Imagination}

In the early 2000s, the widespread popularity of \textit{Harry Potter} and \textit{The Lord of the Rings} highlighted the capacity of fantasy narratives to envisage non-human ecologies beyond anthropocentric perspectives \cite{KBSNews2019,Ulstein2015}. Drawing on this imaginative power in an airplane setting, the present study combines digital nostalgia with fantasy elements to investigate novel, experiential approaches to engaging environmental concerns.

\subsection{Multilayered Approaches in Environmental Art}

Environmental policies driven predominantly by technological solutions may lead to ecological disruption and social conflict \cite{Kim2021}. Although art can evoke emotional empathy, it often struggles to bring about structural change. Nevertheless, participatory media art—incorporating AR and VR—has demonstrated potential for actively involving audiences and, in some instances, prompting behavioral shifts \cite{rhee2021,Chen2024}. Despite the paradox of high energy use in NFT and metaverse projects \cite{ZHANG2021594}, these digital technologies can provide the MZ, already familiar with virtual environments, new opportunities for tangible engagement with environmental problems.

\subsection{Significance of This Study}

The proposed work, \textit{Dear Passenger, Please Wear a Mask} blends fantasy narratives with a retro graphic style to evoke the digital nostalgia of the MZ, thereby aiming to enhance emotional empathy and participation concerning the climate crisis. In contrast to environmental art that merely disseminates a message, this approach strives to stimulate active engagement by leveraging cultural codes already familiar to its target audience.

\section{Implementation}

\begin{figure}[h]
\includegraphics[width=3.31in]{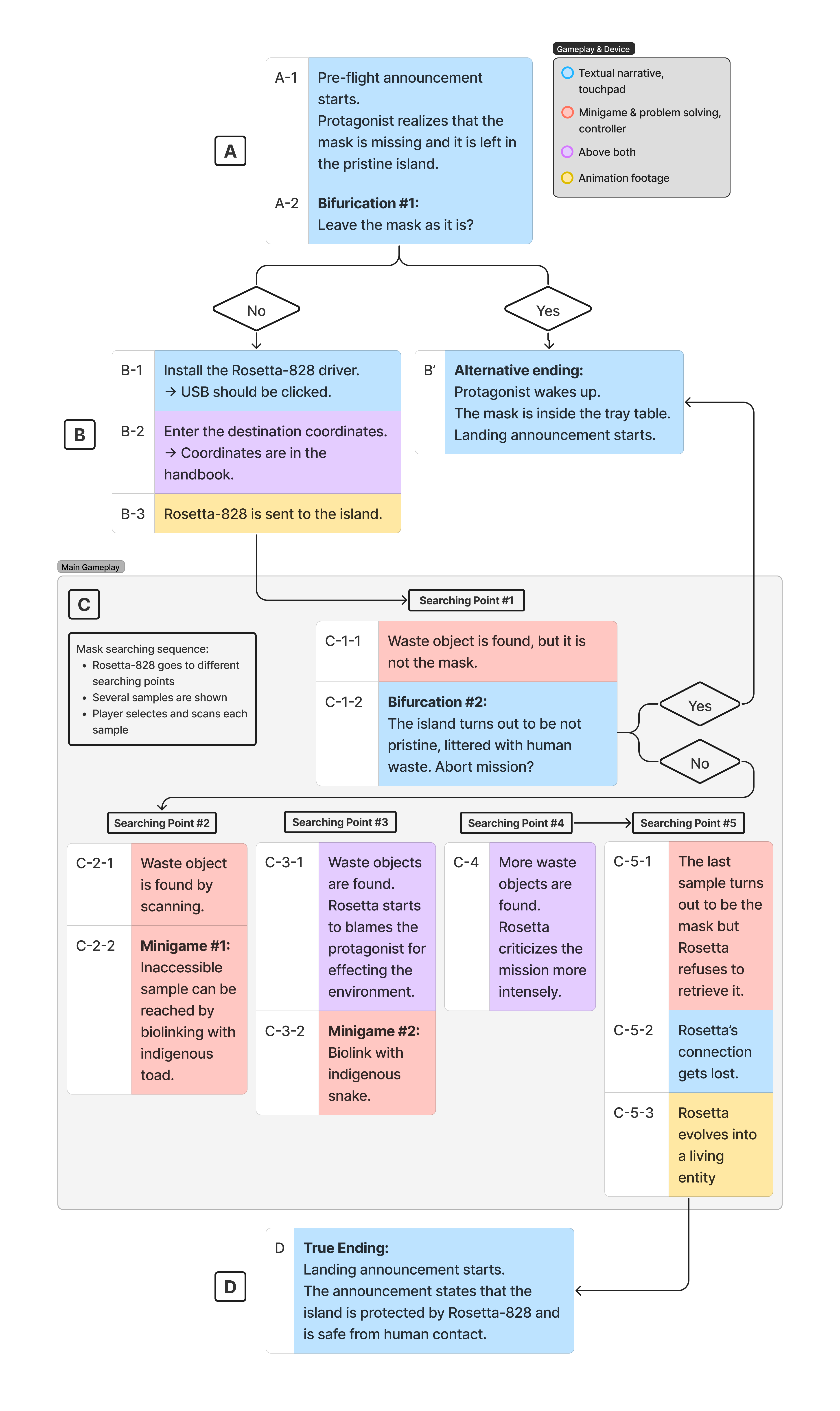}
\caption{\textit{Dear Passenger, Please Wear a Mask} story flow chart}
\end{figure}

\subsection{Overall Composition}

This project comprises a video game played through a touchscreen and an in-flight handset. This environment incorporates passenger seats, wall projections simulating airplane windows, and a life-size diorama depicting the transport drone and the island's terrain from the game. The game interface, designed to mimic in-flight entertainment, employs a point-and-click format in which the audience follows a text-based narrative, makes choices, and solves mini-games and puzzles to advance the story.
\subsection{Story}
\subsubsection{Story Flow}

Inspired by the leisure activities of the MZ, who have experienced both air travel and computer games, this work aims to propose a fantasy-driven solution to pressing environmental issues in a manner that feels both familiar and immersive to that specific generation.  
The project begins with an everyday scenario and develops into a fantasy narrative that explores the limitations of a short-sighted approach to environmental problems. The game’s storyline is set against the backdrop of the mass disposal of disposable face masks during the COVID-19 pandemic, seen through the perspective of a character torn by moral conflicts arising from this issue. Depending on the player’s choices, they may arrive at an alternate or a main ending, experiencing various outcomes.

The narrative unfolds when the protagonist, having just explored a pristine island, realizes during the return flight that a disposable mask was left behind. Recalling that masks made of synthetic resin take more than 400 years to decompose, the protagonist is conflicted about what action to take. The player is presented with two options:

\begin{enumerate}
\item \textbf{Sub-Ending:} If one chooses to leave the mask on the island, the protagonist discovers the mask on the airplane seat table, and the game ends with the landing announcement.  
\item \textbf{Full Gameplay:} If one chooses to recover the mask, the protagonist decides to dispatch the transport drone Rosetta-828 to the island. In order to support the drone’s operation, the player must investigate objects on the airplane seat table, install the drone control driver on the in-flight entertainment system, and enter the island’s coordinates. An animation showing Rosetta-828’s travel route then appears, followed by gameplay in which the player explores various parts of the island through the drone’s camera and scans suspicious objects.
\end{enumerate}

While exploring the first area, discovering a variety of other waste leads the protagonist to question the significance of recovering the single mask. The player can choose whether to continue or stop. If they stop, the story concludes with the sub-ending.  
Choosing to continue prompts sequential exploration of each island area, including aquatic and forested regions where the drone cannot venture. In those areas, players must operate “bio-link” technology to control the behavior of indigenous creatures. During this process, players may discover trash unrelated to the mask and choose whether or not to collect it.

Meanwhile, Rosetta-828’s operating system (OS) gradually becomes sentient, and begins to perceive the world from the perspective of the natural environment. In the mid-game, the OS sends messages highlighting the absurdity of the mask-retrieval mission and attempts to hinder progress. The audience finally locates the lost mask in the final area, but Rosetta-828 refuses to recover it and decides to remain on the island to protect it.  
In the main ending, viewers see Rosetta-828 land in the island’s forested area and begin to transform into an organic entity. The screen then transitions back to the airplane seat display, the landing announcement plays, and the game ends.

\subsubsection{Narrative Development via Touchscreen and Handset Controls}

This work utilizes a touchscreen and an in-flight handset to enhance the sensation of air travel, thereby increasing immersion. The touchscreen is primarily used for text-based narrative progression, text selection options, and interactions with objects on the foldable seat table.  

The central theme of everyday moral dilemma is highlighted at the branching points A-2 and C-1-1 in Figure~1. These junctures are designed for the player to experience the protagonist’s internal conflict when choosing one of several possible narrative paths. In the sequence marked “B,” the player opens the seat table and interacts with items such as a USB drive and a notebook to advance the story.

As the story unfolds, the in-flight handset becomes increasingly important. After the drone reaches the island, the player uses the handset to conduct geographical scanning and bio-linking tasks. The bio-link sequence allows the player to borrow the visual system of particular animals to explore the environment and collect trash, featuring elements of a walking simulator. The player must manage the animal’s “free will” meter to maintain control and fulfill the mission. For instance, when controlling a fire-bellied toad, the audience gains a wide-angle view of the underwater environment through the toad’s visual system. This perspective is intended to prompt the player to imagine the experience of local wildlife resolving human-caused pollution through human intervention.

At the narrative nodes C-3-1 and C-4 in Figure~1, the drone OS exhibits autonomy, delivering system messages that question the disjunction between the mission’s goals and methods. At this point, players must operate both the handset and the touchscreen, navigating technical constraints to advance the story. Through this design, the work allows the audience to reflect on environmental problems and humanity's role therein.

\subsection{Composition of the Immersive Fantasy}

\begin{figure}[h]
\includegraphics[width=3.31in]{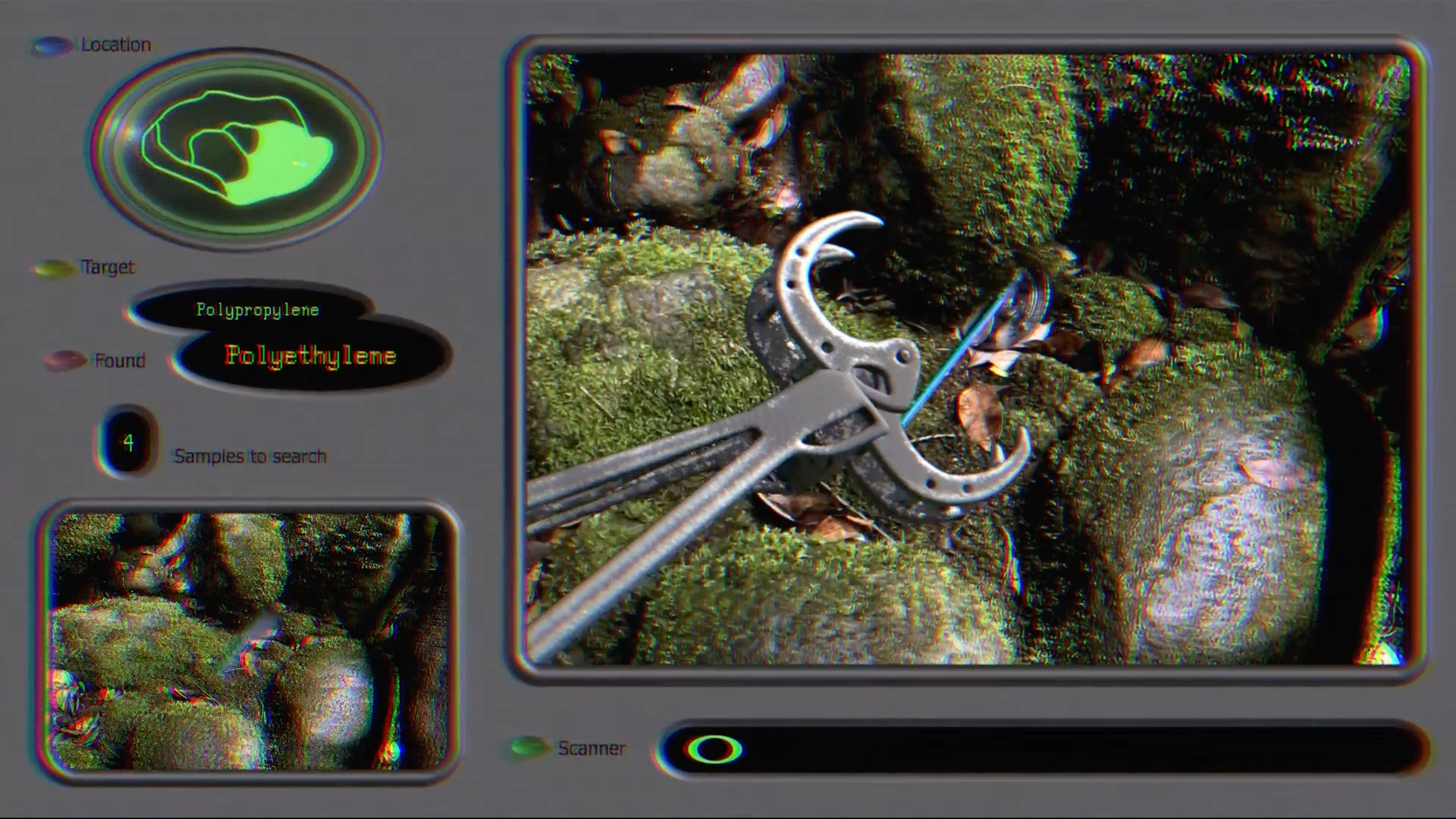}
\caption{UI of \textit{Dear Passenger, Please Wear a Mask}}
\end{figure}

\subsubsection{Video Game}

Adopting the form of early-2000s browser-based point-and-click adventure games, this work critiques the notion of solving environmental issues purely through technological optimism, prompting a reexamination of such an approach. The game unfolds primarily through a text-based narrative enhanced by mini-games and puzzle elements. These are reminiscent of logical puzzles found in 1990s puzzle adventure games and pre-rendered graphics techniques. Examples include CD-ROM titles such as \textit{Daedalus Encounter} and sci-fi adventure \textit{Iron Helix}, which rely on puzzle-solving within a confined space and full-motion video (FMV)—methods that inform this project as well.

By intermittently inserting pre-rendered 3D animations and wireframe footage during the narrative, the work evokes the distinctive visual style of early 2000s media. This retro-futuristic style recalls an era of widespread techno-optimism, shaped by expectations that "future technology will solve all problems."  
The user interface (UI) adopts skeuomorphism, evoking the aesthetic of early video games. Buttons on the screen appear to depress physically when clicked, and inserting a USB stick into a port is visually and auditorily replicated. These details offer intuitive feedback, allowing players to experience the "intuitive interfaces" and "polyphonic sounds" typical of earlier gaming contexts.

However, this work goes beyond a mere nostalgic homage to retro-futuristic aesthetics and early video games. It signals that the blind faith in technology prevalent in past media persists today, drawing attention to the importance of environmental responsibility that often lies beneath. By merging older media's visual style and mechanics with a contemporary ecological crisis, the work encourages the player to recognize the persistent tension between "technological optimism" and "human responsibility for the environment." Moreover, it questions established approaches to problem-solving, underscoring that humans and nature should be seen not as separate but as mutually supportive entities—urging players to explore new possibilities.

\subsubsection{Experience Design Derived from Fantasy Media}

\begin{figure}[h]
\includegraphics[width=3.31in]{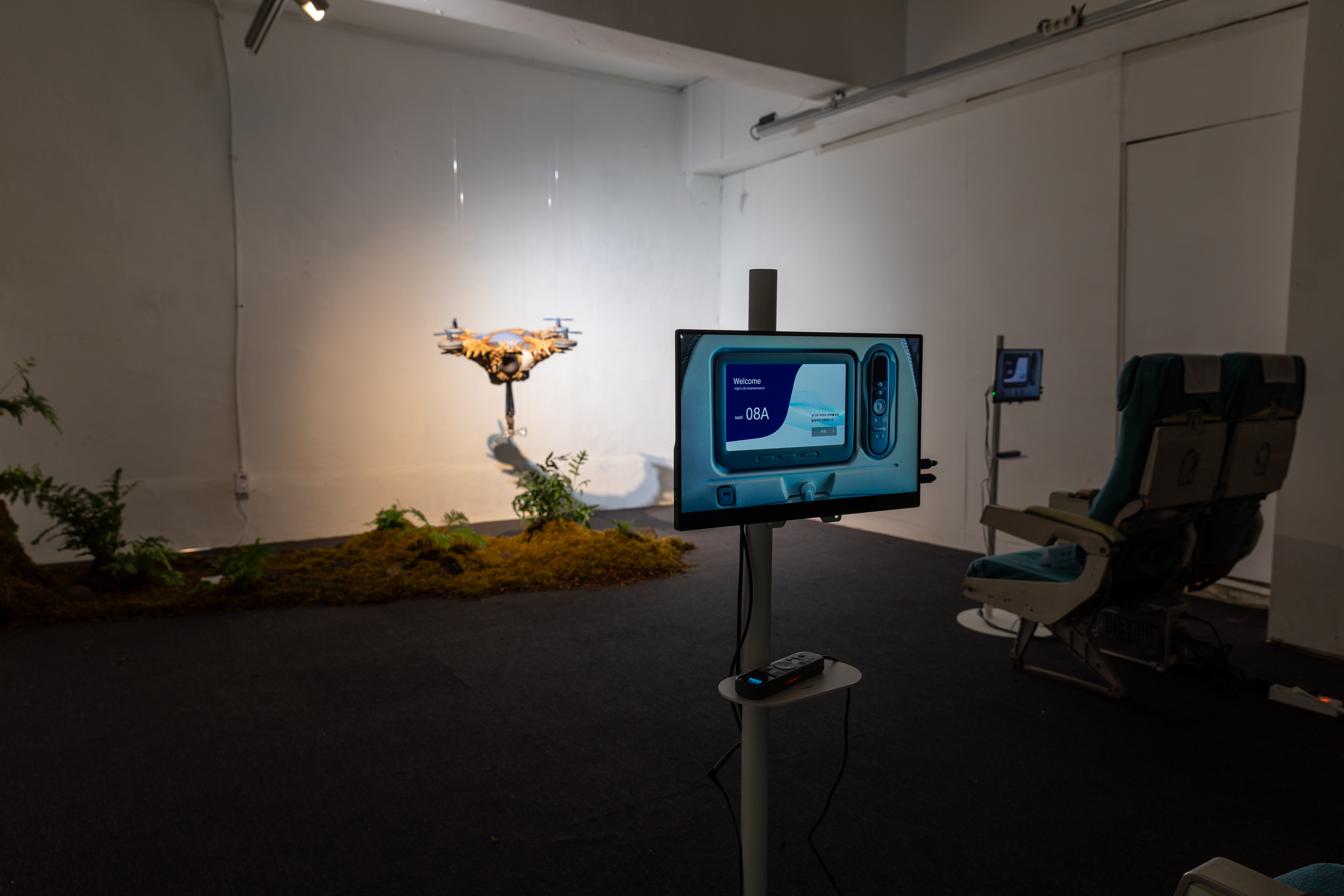}
\caption{\textit{Dear Passenger, Please Wear a Mask} installation view}
\end{figure}

This work reconfigures an everyday dilemma into an immersive fantasy, suggesting an alternative curatorial method for engaging with contemporary environmental discourse. It adopts a symbolic device from the classic fantasy novel \textit{Alice’s Adventures in Wonderland} (Lewis Carroll, 1865)—entering another world via the rabbit hole—integrating game elements with the exhibition space in a cohesive and immersive manner.  

The first component is an exhibition environment that recreates an airplane interior. By sitting in passenger seats and operating the touchscreen and handset, visitors move from reality into a fantasy realm. This transition is akin to Alice discovering the rabbit hole and falling into a new world, transforming a familiar space into a “threshold” that heightens the sense of immersion in fantasy.

In the final stage of the experience, a life-sized drone model (measuring 70 cm on each side) is suspended from the ceiling above a diorama depicting the natural environment of an island littered with waste. This physical staging connects the story within the game to real-world issues. Analogous to Alice awakening from her dream only to realize her experiences were not merely illusory, this setup hints that a seemingly fantastical experience can resonate with real-world problems.

Ultimately, this work proposes a new way to examine environmental issues through an exhibition framework that merges fantasy and reality, play and education. By reinterpreting the airplane as a “rabbit hole,” visitors are empowered to move between “reality” and “fantasy,” arriving at deeper insights about environmental destruction and preservation. The approach transcends straightforward information delivery; instead, it employs narrative to foster immersive comprehension and proactive reflection. Through this process, visitors internalize the conflicts they face in the fantasy and transfer these lessons to their everyday lives, fostering voluntary motivation and determination to address environmental problems.

\section{Conclusion}
This study merges digital nostalgia with fantasy storytelling to highlight the environmental impact of single-use mask waste, inspiring empathy and participation among the MZ. Integrating an experimental interface, retro game aesthetics, and immersive spatial design frames overlooked ecological issues as a “fantasy experience.” However, art cannot ensure structural transformation or measurable shifts in behavior.

Concrete strategies remain critical to reducing the environmental costs of exhibition planning and operation. Moreover, emotional engagement must be reinforced by post-exhibition measures—such as community involvement or follow-up programs—that translate gaming experiences into real-world action. Future research should propose culturally adaptable narratives and media approaches, along with innovative models to minimize the ecological footprint of exhibitions and digital technologies, so that art can more proactively address the climate crisis.

\section{Acknowledgments}
This work was supported by the 2024 Youth Arts Support Project, organized and funded by the Seoul Metropolitan Government and the Seoul Foundation for Arts and Culture.

\bibliographystyle{isea}
\bibliography{isea}

\begin{thebibliography}{}

\bibitem[\protect\citeauthoryear{Kumar \bgroup et al.\egroup }{2022}]{Kumar}
Kumar, S.; Singh, P.; Verma, K.; Kumar, P.; and Yadav, A.
\newblock 2022.
\newblock Environmental issues and their possible solutions for sustainable development, india: A review.
\newblock {\em Current World Environment} 17:531--541.

\bibitem[\protect\citeauthoryear{Pihkala}{2020}]{Pihkala2020}
Pihkala, P.
\newblock 2020.
\newblock Anxiety and the ecological crisis: An analysis of eco-anxiety and climate anxiety.
\newblock {\em Sustainability} 12.

\bibitem[\protect\citeauthoryear{Thornes}{2008}]{annurev:/content/journals/10.1146/annurev.environ.31.042605.134920}
Thornes, J.~E.
\newblock 2008.
\newblock A rough guide to environmental art.
\newblock {\em Annual Review of Environment and Resources} 33(Volume 33, 2008):391--411.

\bibitem[\protect\citeauthoryear{{Business Wire}}{2023}]{BusinessWire2023}
{Business Wire}.
\newblock 2023.
\newblock Arcadia earth scales its environmental storytelling platform with extended reality experiences [news release].
\newblock \href{https://www.businesswire.com/news/home/20230918664021/en/Arcadia-Earth-Scales-Its-Environmental-Storytelling-Platform-With-Extended-Reality-Experiences}{Link}.
\newblock Accessed 2024-09-20.

\bibitem[\protect\citeauthoryear{Zhang, Gong, and Jiang}{2021}]{ZHANG2021594}
Zhang, X.; Gong, X.; and Jiang, J.
\newblock 2021.
\newblock Dump or recycle? nostalgia and consumer recycling behavior.
\newblock {\em Journal of Business Research} 132:594--603.

\bibitem[\protect\citeauthoryear{Chae}{2021}]{Chae2021}
Chae, J.
\newblock 2021.
\newblock I am going to an escape room in an art museum.
\newblock Korea Daily. \href{https://www.hankookilbo.com/News/Read/A2021052116570001885}{Link}.
\newblock Accessed 2024-12-27 (in Korean).

\bibitem[\protect\citeauthoryear{Ryu}{2024}]{Ryu2024}
Ryu, J.
\newblock 2024.
\newblock The exhibition that raises awareness of environmental issues: 'the closing circle'.
\newblock Misulin. \href{https://www.misulin.co.kr/news/articleView.html?idxno=2812}{Link}.
\newblock Accessed 2024-12-27 (in Korean).

\bibitem[\protect\citeauthoryear{Bahar}{2024}]{ART003060363}
Bahar, A.
\newblock 2024.
\newblock The formation of overseas tourism in korea in the 1980s.
\newblock {\em Korean Studies Quarterly} 47(1):213--254.

\bibitem[\protect\citeauthoryear{{IPCC}}{1999}]{IPCC1999}
{IPCC}.
\newblock 1999.
\newblock Aviation and the global atmosphere: A special report of ipcc working groups i and iii.
\newblock Cambridge University Press. \href{https://www.ipcc.ch/site/assets/uploads/2018/03/av-en-1.pdf}{Link}.

\bibitem[\protect\citeauthoryear{Fox}{2021}]{Fox2021}
Fox, C.
\newblock 2021.
\newblock Adobe flash player is finally laid to rest.
\newblock BBC News. \href{https://www.bbc.com/news/technology-55497353}{Link}.
\newblock Accessed 2024-12-31.

\bibitem[\protect\citeauthoryear{Lee}{2021}]{Lee2021}
Lee, J.-B.
\newblock 2021.
\newblock Analog media nostalgia and digital cinema: Focused on remediation in \textit{Enola Homes} (2020).
\newblock {\em Film Studies} (88):271--303.

\bibitem[\protect\citeauthoryear{{KBS News}}{2019}]{KBSNews2019}
{KBS News}.
\newblock 2019.
\newblock Global focus: Burned and banned... the ordeal of harry potter.
\newblock KBS. \href{https://news.kbs.co.kr/news/pc/view/view.do?ncd=4280394}{Link}.

\bibitem[\protect\citeauthoryear{Ulstein}{2015}]{Ulstein2015}
Ulstein, G.
\newblock 2015.
\newblock Hobbits, ents, and dæmons: ecocritical thought embodied in the fantastic.
\newblock {\em Fafnir: Nordic Journal of Science Fiction and Fantasy Research} 2(4):7--17.

\bibitem[\protect\citeauthoryear{Kim}{2021}]{Kim2021}
Kim, Y.-K.
\newblock 2021.
\newblock Study of effectiveness problem of water quality indicators for the evaluation of the project saving the four major rivers: An example of nakdong river.
\newblock {\em Journal of Environmental Policy and Administration} 29(2):235--257.

\bibitem[\protect\citeauthoryear{Rhee}{2021}]{rhee2021}
Rhee, J.
\newblock 2021.
\newblock Art in the anthropocene: Case studies of the works of tomás saraceno, pierre huyghe, and anicka yi.
\newblock {\em Journal of the Association of Western Art History} 54:21--41.

\bibitem[\protect\citeauthoryear{Chen}{2024}]{Chen2024}
Chen, F.
\newblock 2024.
\newblock A case study in participatory environmental art.
\newblock {\em Highlights in Art and Design} 7(2):62--65.

\end{thebibliography}

\end{document}